\documentclass[aps,prd,amsfonts,amssymb,amsmath,nofootinbib,floatfix,reprint,showpacs,groupedaddress]{revtex4-1}
\usepackage{graphicx}
\usepackage{hyperref}

\newcommand{\eqnref}[1]{(\ref{eq:#1})}

\newcommand{\Figref}[1]{Figure~\ref{fig:#1}}
\newcommand{\secref}[1]{Sec.~\ref{sec:#1}}

\newcommand{\units}[1]{\ensuremath{~\mathrm{#1}}}

\newcommand{\sub}[1]{\ensuremath{_\text{#1}}}

\newcommand{\dd}{\ensuremath{\mathrm{d}}}
\newcommand{\diff}[2]{\ensuremath{\frac{\dd {#1}}{\dd {#2}}}}
\newcommand{\intd}[4]{\ensuremath{\int_{#1}^{#2}{#3}\,\dd{#4}}}
\newcommand{\recip}[1]{\ensuremath{\frac{1}{#1}}}

\begin{document}


\title{Gravitational wave energy spectrum of a parabolic encounter}

\author{Christopher P. L. Berry}
\email[]{cplb2@ast.cam.ac.uk}
\author{Jonathan R. Gair}
\email[]{jgair@ast.cam.ac.uk}
\affiliation{Institute of Astronomy, Madingley Road, Cambridge, CB3 0HA, United Kingdom}

\date{\today}

\begin{abstract}
We derive an analytic expression for the energy spectrum  of gravitational waves from a parabolic Keplerian binary by taking the limit of the Peters and Mathews spectrum for eccentric orbits. This demonstrates that the location of the peak of the energy spectrum depends primarily on the orbital periapse rather than the eccentricity. We compare this weak-field result to strong-field calculations and find it is reasonably accurate ($\sim10\%$) provided that the azimuthal and radial orbital frequencies do not differ by more than $\sim10\%$. For equatorial orbits in the Kerr spacetime, this corresponds to periapse radii of $r_{p} \gtrsim 20 M$. These results can be used to model radiation bursts from compact objects on highly eccentric orbits about massive black holes in the local Universe, which could be detected by the Laser Interferometer Space Antenna (LISA).
\end{abstract}


\maketitle

\section{Introduction}

An important source of gravitational waves for the proposed space-based gravitational wave detector, the Laser Interferometer Space Antenna (LISA)~\cite{Bender1998, Danzmann2003}, are the inspirals of stellar-mass compact objects into massive black holes in the centres of galaxies. During the last few years of inspiral these systems generate continuous gravitational waves in the LISA band, which will allow the detection of as many as several hundred systems out to redshift $z \sim 1$~\cite{Gair2009}. However, prior to this phase, the inspiraling object spends many years on a highly eccentric orbit, generating bursts of gravitational radiation at each periapse passage. LISA could resolve individual bursts from sources in the nearby Universe. Initial estimates~\cite{Rubbo2006} suggested a LISA event rate of $\sim18\units{yr^{-1}}$, including $\sim15\units{yr^{-1}}$ from the centre of the Milky Way, although this was subsequently revised downwards to $\sim1\units{yr^{-1}}$ with negligible contribution from extragalactic sources~\cite{Hopman2007}. If even a single burst from the Galactic centre is detected during the LISA mission, this will provide an unparalleled probe of the structure of spacetime there.

The spectrum of radiation from these bursts will be well approximated by the spectrum of a parabolic orbit.\footnote{We use ``parabolic'' to refer to marginally bound orbits. Marginally bound Keplerian orbits are parabola; in curved spacetimes they do not retain such a simple shape.} In this note we derive an analytic approximation to this spectrum by taking the limit of the Peters and Mathews~\cite{Peters1963, Peters1964} (PM) energy spectrum for eccentric Keplerian binaries. We show that the peak of the spectrum depends primarily on the orbital periapse and only weakly on the eccentricity. We also estimate the range of validity of the approximation (in \secref{application}) by comparing to numerical Teukolsky data, finding that it is a good approximation for equatorial orbits in Kerr with periapse $r_{p} \gtrsim 20M$. The parabolic spectrum takes a neat analytic form; deriving it from the bound spectrum will allow corrections for high-eccentricity bound orbits to be found in the future. We hope this note will be a useful resource for future work on gravitational radiation from high-eccentricity orbits.

\section{Parabolic Limit\label{sec:limit}}

\subsection{Energy Spectrum}

For an orbit of eccentricity $e$ with periapse radius $r_{p}$, Peters and Mathews~\cite{Peters1963} give the power radiated into the $n$th harmonic of the orbital angular frequency as
\begin{equation}
P(n) = \frac{32}{5}\frac{G^4}{c^5}\frac{M_1^2M_2^2(M_1 + M_2)(1-e)^5}{r_{p}^5}g(n,e),
\label{eq:PM_P}
\end{equation}
where the function $g(n,e)$ is defined in terms of Bessel functions of the first kind
\begin{eqnarray}
g(n,e) & = & \frac{n^4}{32}\left\{\left[J_{n-2}(ne) - 2eJ_{n-1}(ne) \vphantom{\frac{2}{n}J_n(ne)} \right. \right. \nonumber \\*
 & & \left. + \frac{2}{n}J_n(ne) + 2eJ_{n+1}(ne) - J_{n+2}(ne)\right]^2 \nonumber \\*
 & & + \left(1 - e^2\right)\left[J_{n-2}(ne) - 2J_n(ne) + J_{n+2}(ne)\right]^2 \nonumber \\*
 & & \left. + \frac{4}{3n^2}\left[J_n(ne)\right]^2\right\}.
\end{eqnarray}
The Keplerian orbital frequency is
\begin{equation}
\omega_1^2 = \frac{G(M_1 + M_2)(1 - e)^3}{r_{p}^3} = (1 - e)^3\omega_{c}^2,
\label{eq:Kepler_freq}
\end{equation}
where $\omega_{c}$ is defined as the angular frequency of a circular orbit of radius $r_{p}$. The energy radiated per orbit into the $n$th harmonic, that is, at frequency $\omega_n = n\omega_1$, is
\begin{equation}
E(n) = \frac{2\pi}{\omega_1}P(n);
\label{eq:E(n)}
\end{equation}
as $e \rightarrow 1$ for a parabolic orbit, $\omega_1 \rightarrow 0$ as the orbital period becomes infinite. The energy radiated per orbit is then the total energy radiated. The spacing of harmonics is $\Delta\omega = \omega_1$, giving energy spectrum
\begin{equation}
\left.\diff{E}{\omega}\right|_{\omega_n}\omega_1 = E(n).
\end{equation}
Changing to linear frequency $2\pi f = \omega$,
\begin{eqnarray}
\left.\diff{E}{f}\right|_{f_n} & = & \frac{128\pi^2}{5}\frac{G^3}{c^5}\frac{M_1^2M_2^2}{r_{p}^2}(1-e)^2g(n,e) \\*
 & = & \frac{4\pi^2}{5}\frac{G^3}{c^5}\frac{M_1^2M_2^2}{r_{p}^2}\ell(n,e),
\label{eq:PM_spectrum}
\end{eqnarray}
where the function $\ell(n,e)$ is defined in the last line. For a parabolic orbit, we must take the limit of $\ell(n,e)$ as $e \rightarrow 1$.

We simplify $\ell(n,e)$ using the recurrence formulae (Watson~\cite{Watson1995} 2.12)
\begin{eqnarray}
J_{\nu-1}(z) + J_{\nu+1}(z) & = & \frac{2\nu}{z}J_\nu(z)\\*
J_{\nu-1}(z) - J_{\nu+1}(z) & = & 2J'_\nu(z),\label{eq:J_derivative}
\end{eqnarray}
and eliminate $n$ using
\begin{equation}
n = \frac{\omega_n}{\omega_1} = (1-e)^{-3/2}\tilde{f},
\end{equation}
where $\tilde{f} = \omega_n/\omega_{c} = f_n/f\sub{c}$ is a dimensionless frequency. To find the limit we define two new functions
\begin{equation}
A(\tilde{f}) = \lim_{e\rightarrow 1}\left\{\frac{J_n(ne)}{(1-e)^{1/2}}\right\}; \quad B(\tilde{f}) = \lim_{e\rightarrow 1}\left\{\frac{J'_n(ne)}{1-e}\right\}.
\end{equation}
To give a well-defined energy spectrum, both of these must be finite.

The Bessel function has an integral representation
\begin{equation}
J_\nu(z) = \recip{\pi}\intd{0}{\pi}{\cos(\nu\vartheta - z\sin\vartheta)}{\vartheta};
\end{equation}
we want the limit of this for $\nu \rightarrow \infty$, $z \rightarrow \infty$, with $z \leq \nu$. Using the stationary phase approximation, the dominant contribution to the integral comes from the regime in which the argument of the cosine is approximately zero (Watson~\cite{Watson1995} 8.2, 8.43), for small $\vartheta$:
\begin{eqnarray}
J_\nu(z) & \sim & \recip{\pi}\intd{0}{\pi}{\cos\left(\nu\vartheta - z\vartheta + \frac{z}{6}\vartheta^3\right)}{\vartheta}\\*
 & \sim & \recip{\pi}\intd{0}{\infty}{\cos\left(\nu\vartheta - z\vartheta + \frac{z}{6}\vartheta^3\right)}{\vartheta};
\end{eqnarray}
this last expression is an Airy integral and has a standard form (Watson~\cite{Watson1995} 6.4)
\begin{equation}
\intd{0}{\infty}{\cos(t^3 + xt)}{t} = \frac{\sqrt{x}}{3}K_{1/3}\left(\frac{2x^{3/2}}{3^{3/2}}\right),
\end{equation}
where $K_\nu(z)$ is a modified Bessel function of the second kind. Using this to evaluate the limit gives
\begin{equation}
J_\nu(z) \sim \recip{\pi}\sqrt{\frac{2(\nu - z)}{3z}}K_{1/3}\left(\frac{2^{3/2}}{3}\sqrt{\frac{(\nu -z)^3}{z}}\right).
\label{eq:J_nu}
\end{equation}
For our case,
\begin{equation}
J_n(ne) \sim \recip{\pi}\sqrt{\frac{2}{3}}(1-e)^{1/2}K_{1/3}\left(\frac{2^{3/2}\tilde{f}}{3}\right),
\end{equation}
and the first limiting function is well defined,
\begin{equation}
A(\tilde{f}) = \recip{\pi}\sqrt{\frac{2}{3}}K_{1/3}\left(\frac{2^{3/2}\tilde{f}}{3}\right).
\end{equation}

To find the derivative we combine \eqnref{J_derivative} and \eqnref{J_nu}, and expand to lowest order yielding
\begin{eqnarray}
J'_n(ne) & \sim & -\frac{1}{2\pi}\sqrt{\frac{2}{3}}(1-e)\left[2^{3/2}K'_{1/3}\left(\frac{2^{3/2}\tilde{f}}{3}\right) \right. \nonumber \\*
 & & \left. + \recip{\tilde{f}}K_{1/3}\left(\frac{2^{3/2}\tilde{f}}{3}\right)\right].
\end{eqnarray}
We may re express the derivative using the recurrence formula (Watson~\cite{Watson1995} 3.71)
\begin{equation}
K_{\nu-1}(z) + K_{\nu+1}(z) = -2K'_\nu(z)
\end{equation}
to give
\begin{eqnarray}
J'_n(ne) & \sim & \frac{1-e}{\sqrt{3}\pi}\left[K_{-2/3}\left(\frac{2^{3/2}\tilde{f}}{3}\right) + K_{4/3}\left(\frac{2^{3/2}\tilde{f}}{3}\right) \right. \nonumber \\*
 & & \left. - \recip{\sqrt{2}\tilde{f}}K_{1/3}\left(\frac{2^{3/2}\tilde{f}}{3}\right)\right].
\end{eqnarray}
And so finally we obtain the well-defined
\begin{eqnarray}
B(\tilde{f}) & = & \recip{\sqrt{3}\pi}\left[K_{-2/3}\left(\frac{2^{3/2}\tilde{f}}{3}\right) + K_{4/3}\left(\frac{2^{3/2}\tilde{f}}{3}\right) \right. \nonumber \\*
 & & \left. - \recip{\sqrt{2}\tilde{f}}K_{1/3}\left(\frac{2^{3/2}\tilde{f}}{3}\right)\right].
\end{eqnarray}

\begin{figure}
\includegraphics[width=75mm]{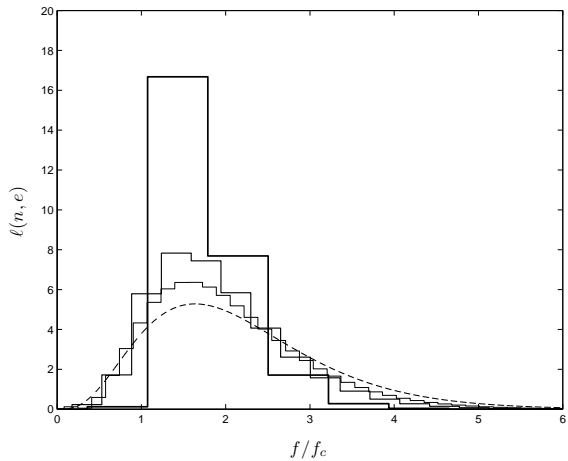}
\caption{The relative energy (per orbit) spectrum $\ell(n,e)$ for $e = 0.2$ (heavy line), $e = 0.5$ (medium line), $e = 0.7$ (light line), and the limiting result for $e = 1$ (dashed line) versus frequency.\label{fig:ell}}
\end{figure}

Having obtained expressions for $A(\tilde{f})$ and $B(\tilde{f})$ in terms of standard functions, we can now calculate the energy spectrum for a parabolic orbit. From \eqnref{PM_spectrum}
\begin{equation}
\diff{E}{f} = \frac{4\pi^2}{5}\frac{G^3}{c^5}\frac{M_1^2M_2^2}{r_{p}^2}\ell\left(\frac{f}{f\sub{c}}\right),
\label{eq:PM_dEdf}
\end{equation}
where we have used the limit
\begin{eqnarray}
\ell(\tilde{f}) & = & \left[8\tilde{f}^2B(\tilde{f}) - 2\tilde{f}A(\tilde{f})\right]^2 \nonumber \\*
 & & + \left(128\tilde{f}^4 + \frac{4\tilde{f}^2}{3}\right)\left[A(\tilde{f})\right]^2.
\end{eqnarray}
This agrees with the $e =1$ form of Turner's result, which was computed by direct integration along unbound orbits~\cite{Turner1977}. \Figref{ell} shows how $\ell(n,e)$ changes with eccentricity including our result for a parabolic encounter (cf. Figure~3 of Peters and Mathews~\cite{Peters1963}). Although more power is radiated into higher harmonics, the peak of the spectrum does not move much: it is always between $f = f_{c}$ and $f = 2 f_{c}$, with $f = 2 f_{c}$ for $e = 0$ and $f \simeq 1.637 f_{c}$ for $e = 1$.

\subsection{Total Energy}

To check the validity of this limit we can calculate the total energy radiated by integrating \eqnref{PM_dEdf} over all frequencies, or by summing the energy radiated into each harmonic. These must yield the same result. Summing:
\begin{equation}
E\sub{sum} = \frac{64\pi}{5}\frac{G^3}{c^5}\frac{M_1^2M_2^2}{r_{p}^2}\omega_{c}(1-e)^{7/2}\sum_n g(n,e),
\end{equation}
where we have used equations \eqnref{PM_P}, \eqnref{Kepler_freq} and \eqnref{E(n)}. Peters and Mathews~\cite{Peters1963} provide the result
\begin{equation}
\sum_n g(n,e) = \frac{1 + (73/24)e^2 + (37/96)e^4}{(1-e^2)^{7/2}}.
\end{equation}
Using this,
\begin{equation}
E\sub{sum} = \frac{64\pi}{5}\frac{G^3}{c^5}\frac{M_1^2M_2^2}{r_{p}^2}\omega_{c}\frac{1 + (73/24)e^2 + (37/96)e^4}{(1+e)^{7/2}},
\end{equation}
which is perfectly well behaved as $e \rightarrow 1$,
\begin{equation}
E\sub{sum} = \frac{85\pi}{2^{5/2}3}\frac{G^3}{c^5}\frac{M_1^2M_2^2}{r_{p}^2}\omega_{c}.
\label{eq:PM_total}
\end{equation}
Integrating the energy spectrum \eqnref{PM_dEdf} gives
\begin{equation}
E\sub{int} = \frac{2\pi}{5}\frac{G^3}{c^5}\frac{M_1^2M_2^2}{r_{p}^2}\omega_{c}\intd{0}{\infty}{\ell(\tilde{f})}{\tilde{f}}.
\end{equation}
The integral can be evaluated numerically as
\begin{equation}
\intd{0}{\infty}{\ell(\tilde{f})}{\tilde{f}} = 12.5216858\ldots = \frac{425}{2^{7/2}3}.
\end{equation}
The two total energies are consistent, $E\sub{int} = E\sub{sum}$.

\section{Applicability\label{sec:application}}

\subsection{Limit of Approximation}

\begin{figure}
\includegraphics[width=75mm]{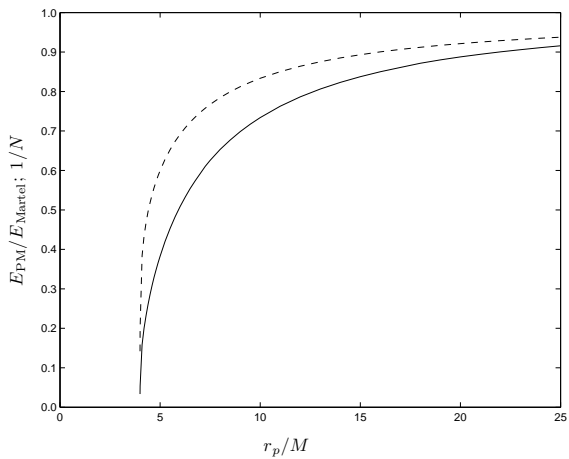}
\caption{Ratio of the total energy radiated as calculated using the Peters and Mathews~\cite{Peters1963} approach to that calculated by Martel~\cite{Martel2004a} using black hole perturbation theory (solid line) versus periapse radius $r_{p}$. The latter approach should give more accurate results. Also shown is the reciprocal of the number of rotations $1/N$ (dashed line). The Keplerian limit corresponds to $N = 1$.\label{fig:Ratio}}
\end{figure}

The PM approach assumes Keplerian orbits in flat spacetime. This should be a valid approximation in the weak-field regime far from a massive body. To find the limit of this approximation, we can compare the PM results with those from more accurate techniques. Energy spectra for parabolic orbits do not seem to be available in the literature yet, so we will make do with the total energy fluxes calculated by Martel~\cite{Martel2004a}, who uses time-domain black hole perturbation theory for a Schwarzschild black hole of mass $M$. \Figref{Ratio} shows the ratio of the two energies as a function of periapse distance. As expected the PM result is more accurate for larger periapses. The agreement worsens as the periapsis decreases. At $r_{p} = 4 M$, corresponding to the radius of the innermost stable circular orbit (ISCO), the energy flux calculated by Martel diverges, so the ratio tends to zero. This divergence is because in Schwarzschild (or Kerr) spacetime a parabolic orbit may have a zoom-whirl structure where it undergoes a number of near circular rotations (whirls) about the black hole. As the radius of the ISCO is approached, the number of whirls tends to infinity (in the absence of radiation reaction), so an infinite amount of energy is radiated.
\Figref{Ratio} also shows how the ratio of energies follows the number of rotations, defined as $N = {\Delta \phi}/{2\pi}$, where $\Delta \phi$ is the total change in the azimuthal angle over one orbit. As $N$ increases, the PM approximation worsens because the Keplerian orbit does not include this extra rotation. The accuracy of the PM result deteriorates rapidly once the orbit transitions to a zoom-whirl trajectory and is therefore far from parabolic in shape.

\begin{figure}
\includegraphics[width=75mm]{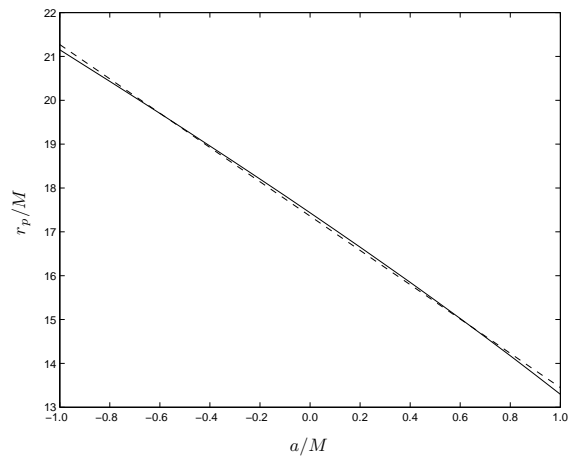}
\caption{Periapse radius corresponding to $N = 1.1$ as a function of spin parameter $a$ (solid line). The curve may be approximated by a straight line $r_{p} = -3.91 a + 17.36M$ (dashed line).\label{fig:N_peri}}
\end{figure}

The PM result is accurate to $\sim 10\%$ for orbits with $N \lesssim 1.1$. We will adopt this as a cutoff point. For an equatorial orbit in Kerr spacetime,
\begin{eqnarray}
N & = & \recip{\pi}\intd{r_{p}}{\infty}{\diff{\phi}{r}}{r} \nonumber \\*
 & = & \frac{L_z}{\pi\sqrt{2M}}\intd{r_{p}}{\infty}{\frac{r^2 - 2M(1 - a/L_z)r}{(r^2 - 2Mr + a^2)w}}{r},
\end{eqnarray}
where
\begin{equation}
w^2 = r^3 - \frac{L_z^2}{2M}r^2 + (L_z - a)^2r;
\end{equation}
$L_z$ is the specific angular momentum about the $z$-axis; $a$ is the spin parameter, and we have adopted units with $G = c = 1$. We will find it useful to define
\begin{equation}
r_\pm = M \pm \sqrt{M^2 - a^2},
\end{equation}
and the two nonzero roots of the cubic $w^2$
\begin{equation}
r_{p,\,1} = \frac{L_z^2}{4M} \pm \sqrt{\frac{L_z^4}{16M^2} - (L_z -a)^2};
\end{equation}
the periapsis is the larger root $r_{p} > r_1$. This equation implicitly gives $L_z$ as a function of $r_p$. The integral may be rewritten as
\begin{equation}
N = \frac{L_z}{\pi\sqrt{2M}}\intd{r_{p}}{\infty}{\recip{w}\left(1 + \frac{\alpha_+}{r-r_+} + \frac{\alpha_-}{r-r_-}\right)}{r},
\end{equation}
where
\begin{equation}
\alpha_\pm = \pm\frac{2Mar_\pm - a^2L_z}{2L_z\sqrt{M^2-a^2}}.
\end{equation}
This may be evaluated using elliptic integrals (Gradshteyn and Ryzhik~\cite{Gradshteyn2000} 3.131.8, 3.137.8)
\begin{equation}
N = \frac{L_z}{\pi}\sqrt{\frac{2}{r_{p}M}}\left[\frac{\alpha_+}{r_+}\Pi\left(\frac{r_+}{r_{p}}\middle|\frac{r_1}{r_{p}}\right) + \frac{\alpha_-}{r_-}\Pi\left(\frac{r_-}{r_{p}}\middle|\frac{r_1}{r_{p}}\right)\right],
\end{equation}
where $\Pi(n|m) = \int_{0}^{\pi/2}{\dd\vartheta/(1-n\sin^2\vartheta)\sqrt{1-m\sin^2\vartheta}}$ is the complete elliptic integral of the third kind. In the limit of $a \rightarrow 0$ we recover the Schwarzschild result~\cite{Cutler1994}
\begin{equation}
N = \frac{L_z}{\pi}\sqrt{\frac{2}{r_{p}M}}K\left(\frac{r_1}{r_{p}}\right),
\end{equation}
where $K(m) = \int_{0}^{\pi/2}{\dd\vartheta/\sqrt{1-m\sin^2\vartheta}}$ is the complete elliptic integral of the first kind. \Figref{N_peri} shows the periapsis for which $N = 1.1$ for a range of spins.
Equatorial orbits with larger periapses should be reasonably approximated by the PM result.

Nonequatorial orbits are more complicated because of the additional precession of the orbital plane. This extra rotation will mean that the PM approach is less accurate; however, this should be subdominant to the perihelion precession effect and so the cutoff periapsis should not be much larger than for the equatorial case.

\subsection{Astrophysical Implications}

Considering bursts from the Galactic centre, orbits with periapses of $r_{p} \lesssim 120 M$ could generate bursts that would be detectable with LISA~\cite{Rubbo2006, Hopman2007}. It is therefore likely that any such burst that was detected would be in the regime of validity of the Peters and Mathews approach, $r_{p} \gtrsim 20 M$ for equatorial orbits. The results described in this note will therefore have application in that context, and it should be possible to explore the majority of parameter space using this approximation. The most interesting orbits, those which come deep within the strong-field region of the black hole's spacetime, will be beyond the range of validity of this approximation, but these represent a small subset of all plausible events.

This result may also be applicable for studying parabolic encounters between stellar mass black holes; these may occur in densely populated environments such as globular clusters~\cite{Kocsis2006} or the Galactic centre~\cite{O'Leary2009}. Bursts from these encounters should be detectable with near-future ground-based detectors, such as the Advanced Laser Interferometric Gravitational-Wave Observatory~\cite{Kocsis2006, O'Leary2009}.

\begin{acknowledgments}
CPLB is supported by STFC. JRG is supported by the Royal Society.
\end{acknowledgments}

\bibliography{Parabolic_spectrum}

\end{document}